\documentclass{sag00}

\begin{document}
\SetRunningHead{Y. Takeda}{Potassium Abundances of Galactic Disk Stars}
\Received{2019/10/16}
\Accepted{2019/12/04}

\title{Determination of Potassium Abundances \\
for Giant and Dwarf Stars in the Galactic Disk}

\author{
Yoichi \textsc{Takeda}\altaffilmark{1,2}
}
\altaffiltext{1}{National Astronomical Observatory, 2-21-1 Osawa, Mitaka, Tokyo 181-8588, Japan}
\email{takeda.yoichi@nao.ac.jp}
\altaffiltext{2}{SOKENDAI, The Graduate University for Advanced Studies, 
2-21-1 Osawa, Mitaka, Tokyo 181-8588, Japan}



%



\maketitle

\begin{abstract}
An extensive study on the potassium abundances of late-type stars was 
carried out by applying the non-LTE spectrum-fitting analysis to the K~{\sc i} 
resonance line at 7698.96~\AA\ to a large sample of 160 FGK dwarfs and 328 
late-G /early-K giants (including 89 giants in the {\it Kepler} field with 
seismologically known ages) belonging to the disk population 
($-1 \ltsim$~[Fe/H]~$\ltsim 0.5$), which may provide important observational 
constraint on the nucleosynthesis history of K in the galactic disk. 
Special attention was paid to clarifying the observed behaviors of [K/Fe] 
in terms of [Fe/H] along with stellar age, and to checking whether giants and dwarfs 
yield consistent results with each other. The following results were obtained.
(1) A slightly increasing tendency of [K/Fe] with a decrease in [Fe/H]
(d[K/Fe]/d[Fe/H]~$\sim -0.1$ to $-0.15$; a shallower slope than reported 
by previous studies) was confirmed for FGK dwarfs, though thick-disk 
stars tend to show larger [K/Fe] deviating from this gradient. 
(2) Almost similar characteristics was observed also for apparently bright field 
giants locating in the solar neighborhood (such as like dwarfs).
(3) However, the [K/Fe] vs. [Fe/H] relation for more distant {\it Kepler} giants
shows larger scatter and is systematically higher (by $\ltsim 0.1$~dex) 
than that of dwarfs, implying that chemical evolution of K is rather diversified 
depending on the position in the Galaxy.   
(4) Regarding the age-dependence, a marginal trend of increasing [K/Fe] with age is 
recognized for dwarfs, while any systematic tendency is not observed for {\it Kepler} giants.
These consequences may suggest that evolution of [K/Fe] with time in the galactic 
disk does exist but proceeded more gradually than previously thought, and its condition
is appreciably location-dependent.

\end{abstract}

\section{Introduction}

While our understanding on the nucleosynthesis history of 
various elements in the Galaxy has remarkably grown
thanks to accumulated observational data of stellar abundances 
as well as progress in theoretical modeling of chemical evolution,
significant discrepancies between theory and observation still remain
for several elements. As one of such cases, it has been known that 
abundance behaviors of potassium (K) observed in old metal-poor 
stars do not well match theoretical predictions (see, e.g., 
Fig.~13 in Prantzos et al. 2018 and the references therein).

For the sake of ameliorating this situation, collecting reliable 
K abundances for as many stars as possible should be important.
However, spectroscopically established abundance data published so far 
are not necessarily sufficient as long as this element is concerned.
This is presumably due to the fact only the strong K~{\sc i} 7698.96~\AA\ 
line of the doublet (hereinafter often referred to as ``K~{\sc i} 7699'') 
has to be invoked, since the paired line at 7664.90~\AA\ (K~{\sc i} 7665) 
tends to be severely blended with telluric lines and other potassium lines 
are generally too weak to be measurable. 
In addition, because of its conspicuous strength, special difficulties are 
involved with abundance determination from this line: one problem is 
the sensitivity to a choice of (often uncertain) damping parameters,
but more important is that it suffers a considerable non-LTE effect.

Actually, pioneering studies of stellar K abundances done under the 
assumption of LTE (e.g., Gratton \& Sneden 1987a,b; Chen et al. 2000) 
are not regarded as reliable as viewed from present-day standard.
Meanwhile, efforts of investigating the non-LTE effect on K~{\sc i} 7699
by non-LTE calculations with sufficiently complex atomic model of K   
have been made by several investigators not only for the Sun 
(de~la~Reza \& M\"{u}ller 1975; Bruls et al. 1992; Zhang et al. 2006a) 
but also for stars of other types (Takeda et al. 1996; Ivanova \& Shimanski\u{\i} 
2000; Takeda et al. 2002; Shimansky et al. 2003; Zhang et al. 2006b; 
Andrievsky et al. 2010; Reggiani et al. 2019). Making use of these theoretical 
accomplishments, chemical evolution studies of potassium based on the non-LTE 
abundances derived from K~{\sc i} 7699 have gradually emerged since the beginning 
of this century, where the targets of those investigations are roughly divided 
into two groups: (i) mildly metal-poor or metal-rich stars of disk population 
($-1 \ltsim$~[Fe/H]~$\ltsim 0.5$) (Takeda et al. 2002; Zhang et al. 2006b;
Wang et al. 2009; Zhao et al. 2016; Reggiani et al. 2019)
and (ii) very metal-poor stars of halo population($-4 \ltsim$~[Fe/H]~$\ltsim -2$)  
(Cayrel et al. 2004; Takeda et al. 2009; Andrievsky et al. 2010; Roederer et al. 2014;
Reggiani et al. 2019).

The focus of this study is placed on the former mildly metal-poor or metal-rich 
stars belonging to the galactic disk population. A problem regarding this group
to be clarified is that the behaviors of [K/Fe]\footnote{
As usual, [X/Y] is the logarithmic abundance ratio of element X to element Y
relative to the Sun; i.e., [X/Y] $\equiv 
(A^{\rm X}_{\rm star} -A^{\rm X}_{\odot}) - (A^{\rm Y}_{\rm star} -A^{\rm Y}_{\odot})$,
where $A^{\rm X}$ is the logarithmic number abundance of element X
normalized with $A^{\rm H} = 12$ (the suffix $\odot$ denotes the solar value). 
} vs.[Fe/H] relation (an important diagram for investigating the nucleosynthesis
evolution of K in the Galaxy) derived for disk stars by several
previous studies are not necessarily consistent with each other in the quantitative 
sense. Takeda et al.'s (2002) non-LTE reanalysis on Chen et al.'s (2000) spectra of 
thin-disk stars resulted in a progressive increase of [K/Fe] with a decrease in [Fe/H]
from [K/Fe]~$\sim -0.1$ (at [Fe/H]~$\sim 0$) to [K/Fe]~$\sim$~+0.3 (at [Fe/H]~$\sim -1$)
(cf. Fig.~4a therein), suggesting d[K/Fe]/d[Fe/H]~$\sim -0.4$. 
Meanwhile, Zhang et al.'s (2006b; cf. Fig.~5 therein) results for thin-disk stars
($-0.8 \ltsim$~[Fe/H]~$\ltsim -0.2$) as well as Wang et al.'s (2009; cf. Fig.~9c therein) 
analysis on thin-disk stars ($-0.5 \ltsim$~[Fe/H]~$\ltsim +0.4$) showed a similar 
linear tendency but appreciably steeper gradient (d[K/Fe]/d[Fe/H]~$\sim -0.6$). 
In contrast, Zhao et al.'s (2016) Fig.~9 suggests a shallower slope of 
d[K/Fe]/d[Fe/H]~$\sim -0.2$ to $-0.3$ for thin-disk stars ($-0.8 \ltsim$~[Fe/H]~$\ltsim +0.2$).
Which of these represents the actual trend of disk stars at all?

Here, the following points may be worth consideration:\\
--- First, the number of stars analyzed by each of these studies is only 
around $\sim 20$, which is not sufficient for elucidating statistically 
meaningful tendency. Considerably increasing the sample size (ca., up to 
the order of $\sim 10^{2}$) would undoubtedly improve the situation.\\
--- Second, all these previous K abundance studies for disk-population stars  
made use of late F--early G dwarfs, but red giants (of disk metallicity)\footnote{
In contrast, studies of K abundances in very metal-poor halo stars predominantly 
invoked evolved giants (e.g., Cayrel et al. 2004); thus the situation is
markedly different from the case of galactic disk stars.} 
seem to have never been employed for this purpose. 
Since potassium is not expected to suffer any change by evolution-induced 
dredge-up (unlike lighter elements such as CNO or Na), primordial K abundances 
should be retained at the surface of giants. Then, why not use red giants
(which are intrinsically bright and available in large numbers) for 
investigating the chemical evolution of K in the disk of the Galaxy? \\
--- Third, since [Fe/H] is roughly related to stellar age (i.e., older stars
tend to have lower [Fe/H]), it is worthwhile to examine age-dependence of [K/Fe]
if possible, since progressive increase with age may as well be detected
if [K/Fe] grows with a lowering of [Fe/H]. In this context, it is interesting
that recent Spina et al.'s (2016) high-precision differential abundance study 
on the solar twins of widely different ages showed that [K/Fe] is almost
age-independent (unlike other elements which revealed systematic trend with age).
The successsive non-LTE reanalysis of Spina et al.'s (2016) data by 
Reggiani et al. (2019) again resulted in almost the same conclusion.    
As such, this point should be worth reinvestigation based on a large 
number of stars with known ages. 

Motivated by these notions, this study aims at determining the
non-LTE K abundances from the K~{\sc i} 7699 line for an extensive sample 
of 160 late F--early K dwarfs as well as 328 late G--early K giants 
(239 nearby giants and 89 giants in the {\it Kepler} field) belonging 
to the disk population ($-1 \ltsim$~[Fe/H]~$\ltsim 0.5$), for which 
the spectra are conveniently available to the author and the stellar
parameters are already determined (especially, stellar ages are
known for FGK dwarfs and {\it Kepler} giants).
The points of interest to be checked upon are as follows:
\begin{itemize}
\item
How are the [K/Fe] vs. [Fe/H] relations derived for each of the groups
(dwarfs, nearby giants, and {\it Kepler} giants) compared with each other? 
Are they consistent, or any difference is seen?
\item
What about the slope (d[K/Fe]/d[Fe/H]) of the resulting [K/Fe] values 
in comparison with the diversified gradients reported by previous studies? 
\item
Is any systematic trend observed in [K/Fe] with stellar age? 
\end{itemize}

\section{Observational Data and Stellar Parameters}

All the observational materials and stellar parameters employed for this 
investigation are those already used and determined in the previous
studies of the author's group.

\subsection{FGK dwarfs}

Regarding dwarfs (including subgiants), the spectra of 160 stars observed 
in 2000--2003 with the 1.9~m reflector (+HIDES spectrograph) at Okayama 
Astrophysical Observatory (OAO) and published by Takeda et al. (2005a) were used. 
Besides, the Moon spectrum (substitute for the reference solar spectrum) was also 
taken from this database. The atmospheric parameters
[$T_{\rm eff}$ (effective temperature), $\log g$ (surface gravity), $v_{\rm t}$
(microturbulence), and [Fe/H] (Fe abundance relative to the Sun)] of these
160 stars were determined from the equivalent widths measured for a number of 
selected Fe~{\sc i} and Fe~{\sc ii} lines, while requiring that abundances
do not depend upon $\chi$ (lower excitation potential) as well as $W$
(equivalent width) and that mean abundances derived from Fe~{\sc i} and 
Fe~{\sc ii} lines are equal (cf. Takeda et al. 2005b).
Regarding the atmospheric parameters of Sun, $T_{\rm eff,\odot} = 5780$~K, 
$\log g_{\odot} = 4.44$, $v_{\rm t,\odot} = 1.0$~km~s$^{-1}$, and [Fe/H]$_{\odot} = 0.00$
were employed (note that practically the same values were empirically derived with this 
method using Fe~{\sc i}/Fe~{\sc ii} lines applied the Moon spectrum; cf. Takeda et al. 2005b).
Takeda (2007) evaluated the age ($\log age$) and mass ($M$) for each of these 
160 stars by comparing the position in the $\log L$ ($L$: bolometric luminosity) 
vs. $\log T_{\rm eff}$ diagram with theoretical stellar evolutionary tracks.
These stellar parameters are summarized in tableE1.dat (online material).

\subsection{Nearby giants}

As to giants, the spectra of 239 late G or early K giants were used, which were 
observed in 2012--2013 with the 1.9~m telescope and HIDES spectrograph at OAO 
and described in Takeda et al. (2015). These 239 stars (apparently bright with
$V < 6$~mag and comparatively near to us within several hundreds parsec)  
are the subset of 322 giants targeted by the Okayama Planet Search Program. 
for which the atmospheric parameters ($T_{\rm eff}$, $\log g$, $v_{\rm t}$, and 
[Fe/H]) were already established  by Takeda et al. (2008) in the same way as done 
for 160 dwarfs mentioned above.
However, the mass values (closely related to stellar age) of these giants estimated 
by Takeda et al. (2008) are likely to be appreciably overestimated and unreliable 
for a number of red clump stars (constituting the majority of these sample stars), 
because of the difficulty in discriminating the evolutionary status of each star 
by using such a coarse grid of theoretical stellar evolutionary tracks as applied 
in that paper (cf. Takeda \& Tajitsu 2015; Takeda et al. 2016 for more details). 
Accordingly, mass and age are treated as undetermined and not mentioned in this 
paper for these 239 giants, for which only four atmospheric parameters are presented 
in tableE2.dat (online material).

\subsection{{\it Kepler} giants}

In addition, as another set of giant stars, the spectra of late G--early K giants 
in the {\it Kepler} field observed  with the 8.2~m Subaru Telescope (+ HDS spectrograph) on 
2014 September 9 (42 stars) and 2015 July 3 (47 stars)\footnote{Actually, 48 stars in the 
{\it Kepler} field were observed in this observing run. However, for the case of KIC~7341231
which is a very metal poor subgiant with especially large (negative) radial velocity 
($V_{\rm r}^{\rm hel} \simeq -270$~km~s$^{-1}$), the K~{\sc i} 7699 line unfortunately 
fell on the gap of the echelle format at 7687--7694~\AA\ and could not be used.
Therefore, this star had to be excluded from this study.} were also employed. 
Their atmospheric parameters were already determined spectroscopically by Takeda \& Tajitsu (2015) 
and Takeda et al. (2016) in same manner as the cases of dwarfs and nearby giants described above.
The distinct merit of these {\it Kepler} field giants is that the stellar mass ($M$) 
as well as the evolutionary status of each star are confidently established by making use 
of the seismological information, by which the stellar age can be reliably determined 
with the help of stellar evolutionary tracks as done by Takeda et al. (2016).
In tableE3.dat (online material) are given the values of $T_{\rm eff}$, $\log g$, 
$v_{\rm t}$, [Fe/H], $M$, and $\log age$ for these 89 giant stars. 

\subsection{Comparison between the groups}

The mutual comparisons of $T_{\rm eff}$, $\log g$, $v_{\rm t}$, and [Fe/H] are 
depicted in Figures~1a--1d, in order to illustrate the difference or similarity 
of these star groups. The discussions regarding correlations between these 
parameters are presented in Takeda et al. (2005b) and Takeda et al. (2008) 
for the case of dwarfs and giants, respectively.  

Figures~1e and 1f also show the positional distributions of the program stars
in the $XYZ$-space, where $X \equiv d \cos b \cos l$, $Y \equiv d \cos b \sin l$, 
and $Z \equiv d \sin b$ ($d$: distance from the Sun, $l$: galactic longitude,
and $b$: galactic latitude). It can be seen from these figures that, while 160 dwarfs
(within several tens parsec) as well as 239 nearby giants (within several hundreds parsec) are 
solar neighborhood stars as viewed from the galactic scale, 89 {\it Kepler} giants 
widely distribute with diversified distances from $\sim 10^{2}$~pc up to $\sim 2$~kpc 
in the confined direction to Cygnus.
Accordingly, from the viewpoint of location in the Galaxy. the former two groups 
(dwarfs, nearby giants) are almost similar but the last group ({\it Kepler} giants) 
is distinctly different. 

\section{Kinematic Properties}

\subsection{Space velocities and orbital parameters}

Examining the kinematic properties of stars is important for understanding to which
stellar population they belong. Takeda (2007) computed the orbital motions 
within the galactic gravitational potential for the 160 FGK stars (program stars of
this study) by using the positional data and the velocity data (proper motions 
and radial velocity), and derived $R_{\rm m}$ (mean galactocentric distance), 
$e$ (orbital eccentricity), $z_{\rm max}$ (maximum separation from the galactic plane).
These orbital quantities along with the space velocity components relative to the Local 
Standard of Rest ($U_{\rm LSR}$, $V_{\rm LSR}$, $W_{\rm LSR}$) are presented in 
electronic tableE2 (kinepara.dat) of that paper.

Likewise, Takeda et al. (2008) carried out similar analysis of kinematic properties 
for 322 red giants (from which 239 nearby giants for this study were selected),
and the resulting parameters are given in electronic tableE1 (kinepara.dat) 
of that paper.

These kinematic parameters for 89 giants in the {\it Kepler} field were newly computed
for this investigation, following the same way as done in Takeda (2007).
Regarding the parallax and proper motion data, the {\it Gaia} DR2 database
(Gaia Collaboration et al. 2016, 2018) was invoked, while the values 
measured from our Subaru HDS spectra were used for radial velocities 
(they were confirmed to be consistent with {\it Gaia} DR2 data).
The resulting orbital parameters and space velocity components are given
in tableE4.dat of the online material.

\subsection{Classification of stellar population}

According to Ibukiyama \& Arimoto (2002), the $z_{\rm max}$ vs. $V_{\rm LSR}$ 
(rotation velocity component) diagram can be used for discriminating thin-disk, 
thick-disk, and halo populations, as already done by Takeda (2007) and Takeda 
et al. (2008) to examine the stellar populations of FGK dwarfs and nearby giants. 
Figure~2a (reproduction of Fig.~14d in Takeda 2007), Figure~2b (similar to
Fig.~9a in Takeda et al. 2008), and Figure~2c illustrate this situation for
160 FGK dwarfs, 239 nearby giants, and 89 {\it Kepler} giants, respectively.    
From these figures, 6 (out of 160) dwarfs, 5 (out of 239) nearby giants,
and 13 (out of 89) {\it Kepler} field giants were concluded to be of thick-disk 
population (indicated by open symbols in the figures; see ReadMe.txt in online 
material for their specific IDs), while all others are of thin-disk population. 

\subsection{Relation to metallicity and age}

Figures~2d--2l also depict the correlations between the space velocity 
$|v_{\rm LSR}|$ ($\equiv \sqrt{U_{\rm LSR}^{2}+ V_{\rm LSR}^{2}+W_{\rm LSR}^{2}}$), 
metallicity ([Fe/H]), and stellar age.
It can be read from these figures that (1) thick-disk stars tend to be old,
metal-poor, and moving faster as compared to thin-disk stars, (2) lower-metallicity
stars as well as older stars are apt to have higher space velocity, and (3) 
there is a tendency of older stars having lower metallicity; all these trends
are natural and reasonably understood. 

\section{Determination of K abundances}

\subsection{Spectrum-fitting analysis}

Although most of the previous studies mentioned in Section~1 determined the potassium 
abundances from the equivalent widths ($W$) of K~{\sc i} 7699 line, precisely measuring 
$W$ is not necessarily an easy task for such a strong line often showing appreciable 
damping wings. In this investigation was instead adopted a spectrum-fitting approach 
for establishing the non-LTE K abundances of the program stars.

The procedure is similar to the one adopted in Takeda et al. (1996), which 
accomplishes the best fit between theoretical and observed profiles by finding 
the most optimum solutions of three parameters [$A$(K) (K abundance)\footnote{
As usual, $A$(X) is the logarithmic number abundance of element X normalized as 
$A({\rm H})=12$: i.e., $A({\rm X}) = \log [N({\rm X})/N({\rm H})] + 12.$}, 
$v_{\rm M}$ (macrobroadening parameter),\footnote{This $v_{\rm M}$ is the $e$-folding
half width of the Gaussian macrobroadening function $f_{\rm M}(v) \propto \exp[{-(v/v_{\rm M})^2}]$,
which is regarded as a combination of (i) instrumental broadening, (ii) macroturbulence 
broadening, and (iii) rotational broadening.} and $\Delta \lambda$ (wavelength shift)], 
while applying the numerical algorithm described in Takeda (1995).

That is, unbroadened non-LTE theoretical profile, which was calculated for the relevant 
model atmosphere and non-LTE departure coefficients (specified by $T_{\rm eff}$, $\log g$, 
and [Fe/H]) along with the given microturbulence ($v_{\rm t}$), was broadened (by convolving 
the broadening function corresponding to $v_{\rm M}$) and shifted (by $\Delta\lambda$) 
in order to be compared with the observed profile.
The model atmosphere of each star was generated by interpolating Kurucz's (1993)
ATLAS9 model grid in terms of $T_{\rm eff}$, $\log g$, and [Fe/H], as done in Takeda (2007)
or Takeda et al. (2008). Similarly, the depth-dependent non-LTE departure coefficients 
of the relevant levels (i.e., ground level and first excited level) were derived by 
interpolating the extensive grid computed in Takeda et al. (2002; cf. Appendix therein). 
Regarding the atomic parameters of the K~{\sc i} 7699 line, the same values as used 
in Takeda et al. (2002) were adopted (cf. Sect.~4 therein).  
Appendix shows that Takeda et al.'s (2002) non-LTE calculations,
upon which ths study is based, are in reasonable agreement with 
the recent ones of Reggiani et al. (2019).

This method turned out successful and a satisfactory fit could be accomplished for 
almost all cases, as demonstrated in Figure~3 (160 FGK dwarfs + Moon), Figure~4
(239 nearby giants) and Figure~5 (89 {\it Kepler} giants).
Note that, since telluric lines (or spectrum defects) sometimes invaded into the
fitting range depending on the radial velocity, they had to be masked and neglected
in judging the goodness of fitting (as highlighted in green in these figures). 

\subsection{Equivalent widths and non-LTE corrections}

Although the non-LTE K abundance ($A$) of each star could be established based on
the spectrum-fitting technique as such, only this information would not suffice.
In order for understanding the behaviors of resulting abundances (e.g., sensitivity to 
stellar parameters) differing from star to star, abundance-related quantities such as 
equivalent width ($W$) or non-LTE correction ($\Delta$) are quite useful.
Therefore, by using Kurucz's (1993) WIDTH9 program which was considerably modified
by the author (e.g., to enable inclusion of non-LTE effect). the corresponding $W$ 
of K~{\sc i} 7699 was evaluated by using the solution of $A$ along with the 
same model atmosphere and $v_{\rm t}$. Furthermore, based on such calculated $W$,
the non-LTE abundance ($A^{\rm NLTE}$) and LTE abundance ($A^{\rm LTE}$) were inversely 
derived, from which the non-LTE abundance correction ($\Delta$) was obtained as 
$A^{\rm NLTE} - A^{\rm LTE}$.  
The resulting $W$, $A^{\rm NLTE}$, and $\Delta$ values are presented in tableE1.dat
(FGK dwarfs), tableE2.dat (nearby giants), and tableE3.dat ({\it Kepler} giants)
of the online material.

Figure~6 illustrates how $W$ and $\Delta$ depend on $T_{\rm eff}$ and $\log g$
and how they correlate with each other. 
Roughly speaking, $W$/$|\Delta|$ tends to decrease/increase with higher $T_{\rm eff}$ 
or lower $\log g$ (cf. Figures~6a--6d), as long as stars belonging to the same group 
(e.g., `dwarfs' group or `giants' group) are concerned.
According to Figure~6e, the extent of $\Delta$ tends to diminish as $W$ increases.
This is due to the fact that comparatively weaker lines 
(100~m\AA~$\ltsim W \ltsim 200$~m\AA) are most saturated (i.e., line-forming
layer is high) at the shoulder or the flat part of curve of growth and suffer 
an especially large non-LTE effect, while those stronger lines 
(200~m\AA~$\ltsim W \ltsim 300$~m\AA) are already in the damping part and 
comparatively less sensitive to a non-LTE effect (because deep-forming damping 
wings appreciably contribute to the total strength of such a strong line).  
This situation is elucidated in Figure~6f, where $\log W$ is plotted against
$A^{\rm NLTE}$ + $(\chi_{\rm ion} - \chi_{\rm exc})5040/T_{\rm eff}$
($\chi_{\rm ion}$ = 4.34~eV, $\chi_{\rm exc}$ = 0.00~eV), which may be 
regarded as the abscissa of curve of growth (controlling the $A$- and 
$T_{\rm eff}$-dependence of line strength).

\section{Behaviors of Potassium Abundances}

\subsection{[K/Fe] vs. [Fe/H] relation}

The [K/Fe] ratios of the program stars were calculated from the potassium abundances 
determined in Section~4 as
\begin{equation}
[{\rm K}/{\rm Fe}] \equiv (A^{\rm NLTE} - A^{\rm NLTE}_{\odot}) - [{\rm Fe}/{\rm H}], 
\end{equation} 
where $A^{\rm NLTE}_{\odot}$ is the solar potassium abundance of 4.96
derived from the spectrum of Moon.

Takeda et al. (2002) discussed the ambiguities involved with the potassium abundances 
they derived from K~{\sc i} 7699 caused by uncertainties in atmospheric parameters
(see Sect.~5 therein). In this investigation, the adopted values of $T_{\rm eff}$, 
$\log g$, $v_{\rm t}$, and [Fe/H] of all program stars are those consistently determined 
by the same method using Fe lines, and their statistical errors are typically several tens 
of K, $\ltsim$~0.1~dex, $\sim$~0.1~km~s$^{-1}$, and a few hundredths of dex, respectively 
(cf. Takeda et al. 2005b, 2008).
By combining these uncertainties with the results of Takeda et al. (2002),
the typical errors in [K/Fe] may be estimated as $\ltsim 0.1$~dex.

The resulting [K/Fe] vs. [Fe/H] diagrams for each of the three stellar groups are displayed
in Figures~7a--7c, where the corresponding non-LTE corrections are also shown.
The characteristic trends recognized from these figures are summarized below:
\begin{itemize}
\item
Generally, a qualitative trend is seen that [K/Fe] tends to increases with a decrease 
in [Fe/H], which is commonly seen in each of the groups (though specific details are
different).
\item
Almost all thick-disk stars (indicated by open symbols in the figures) have supersolar 
[K/Fe] ($>0$), mainly because they tend to be appreciably metal-poor ([Fe/H]~$< 0$).
\item
Regarding the group of dwarf stars, [K/Fe] vs. [Fe/H] relation shows a tight tendency 
with small dispersion (cf. Figure~7a). Only one exceptional outlier is 
HD~50554, which exhibits a very large [K/Fe] of +0.74 despite its near-solar 
metallicity ([Fe/H] = $-0.03$); but the reason for this anomaly is unclear. 
\item
Nearby giants reveal a rather similar trend to the case of dwarfs, though the 
dispersion is apparently larger and the slope appears to be slightly steeper, 
while several outliers with appreciably large [K/Fe] ($>0$) exist even around 
the solar metallicity (cf. Figure~7b).  
\item
However, a different feature is recognized for the case of giants in 
the {\it Kepler} field (cf. Figure~7c). Here, while the increasing trend of [K/Fe]
towards lowered metallicity is still roughly seen by the existence of thick-disk 
stars, [K/Fe] values are generally enhanced (i.e., tending to be supersolar irrespective 
of the metallicity) with a considerable scatter. This implies that the chemical evolution
history of these {\it Kepler} giants would have been dissimilar to that of 
solar neighborhood stars. The reason why they show so diversified [K/Fe] may be
due to their widely scattered locations along the direction to Cygnus 
(cf. Figures~1e and 1f).
\end{itemize} 

\subsection{Comparison between dwarfs and giants}

It is not easy to evaluate the quantitative characteristics 
of [K/Fe] (e.g., gradient with respect to a change in [Fe/H]) in 
Figures~7a--7c only by eye-inspection, because the plotted points are 
densely overlapped. Therefore, mean $\langle$[K/Fe]$\rangle$ (and
the corresponding standard deviation) were computed by averaging the 
[K/Fe] values\footnote{In the averaging process, the data showing 
appreciable deviations within each bin-group were discarded, 
which were judged by Chauvenet's criterion (Taylor 1997).} included 
at each of the seven metallicity bins 
($\Delta$[Fe/H] = 0.1dex at $-0.5 \le$~[Fe/H]~$\le +0.2$) for three 
star groups, and comparisons were made for nearby giants vs. FGK stars
as well as {\it Kepler giants} vs. FGK dwarfs, as shown in Figures~8a and 8b.
The linear-regression analysis applied to these $\langle$[K/Fe]$\rangle$ 
data yielded the following relations:
$\langle$[K/Fe]$\rangle$ = $-0.129 (\pm {\it 0.022})\times$[Fe/H]$-0.008 (\pm {\it 0.005})$ (FGK dwarfs),
$\langle$[K/Fe]$\rangle$ = $-0.157 (\pm {\it 0.013})\times$[Fe/H]$+0.004 (\pm {\it 0.003})$ (nearby giants), and
$\langle$[K/Fe]$\rangle$ = $-0.100 (\pm {\it 0.063})\times$[Fe/H]$+0.049 (\pm {\it 0.016})$ ({\it Kepler} giants),
where the values in parentheses (denoted in Italics) are the standard errors.
These lines are also drawn in the figures.

Figure~8a suggests that the [K/Fe] trend of nearby giants is not much 
different from that of FGK dwarfs (though marginally higher [K/Fe] 
by a few hundredths dex and slightly steeper slope), while {\it Kepler} 
giants and dwarfs (shown in Figure 8b) are in disagreement with each other 
as already mentioned (the former show systematically larger [K/Fe] 
values at the same metallicity).
Consequently, it can be concluded that (1) nearly consistent [K/Fe] vs. [Fe/H] 
relations could be derived from this investigation for both dwarfs and 
giants as long as stars in the solar neighborhood are concerned, (2) 
but the giants in the {\it Kepler} field are different presumably 
because they are in diversified locations far from the Sun where
chemical evolution history would have been more or less variant. 

According to Figure~8a, the d[K/Fe]/d[Fe/H] gradient corresponding to
galactic disk dwarfs in the solar neighborhood (within several tens parsecs)
is concluded to be $\sim -0.1$ to $-0.15$, while that for nearby giants 
(within several hundreds parsec) is similar around $\sim -0.15$.
Regarding the questions raised in Section~1 (i.e., which of the variously 
reported slopes are correct), the resulting d[K/Fe]/d[Fe/H] gradient 
for thin-disk population derived from this investigation (from $\sim -0.1$ 
to $-0.15$) is shallower than any of the past studies,
though the recent Zhao et al.'s (2016) consequence ($\sim -0.2$ to $-0.3$) 
is reasonably the closest among these. That these previous authors obtained 
diversified results of comparatively steeper slope (from $\sim -0.2$ to 
$-0.6$) may have originated from the paucity of their sample stars.

\subsection{Relation to stellar ages}

Finally, how the resulting [K/Fe] depends on the stellar age is examined,
which is possible for FGK dwarfs and {\it Kepler} giants (ages are not available 
for nearby giants; cf. Section~2.2). An inspection of Figure~7d reveals 
that a systematic but subtle increase exists in [K/Fe] with age for these 
dwarfs in the solar neighborhood, which means that [K/Fe] in the gas of
the solar neighborhood gradually evolved from $\sim +0.1$ ($\sim 10^{10}$~yr ago)
to [K/Fe]~$\sim 0$ ($\sim 5\times 10^{9}$~yr ago when the Sun was born)
and further down to $\sim -0.1$ ($\sim 10^{9}$~yr ago).
In contrast, any clear age-dependence is hardly observed in the [K/Fe] values
of {\it Kepler} giants (cf. Figure~7f), which is presumably because stars having
experienced diversified chemical enrichment histories are mixed up in this sample.  

As mentioned in Section~1, precisely determined [K/Fe] values for solar analog 
stars (with various stellar ages from $\sim 10^{9}$~yr to $\sim 10^{10}$~yr) 
determined by Spina et al. (2016) and Reggiani et al. (2019) (non-LTE reanalysis) 
did not show any clear dependence upon age, although most of their sample stars
are in the solar neighborhood. It should be noted, however, that all those solar 
twins have almost the same metallicity as that of the Sun ([Fe/H]~$\simeq 0$), despite 
that they have diversified ages (i.e., metallicity is not dependent upon age at all).
Accordingly, since the target stars used in these two recent studies are rather 
specific, their results can not be directly compared with that of the present 
investigation based on a large number of nearby dwarfs showing a rough
age--metallicity relation (cf. Figure~2j). As a possibility, it may be that 
the enrichment of K in the Galaxy significantly depends upon the local condition 
(i.e., not necessarily a unique function of time) and that [K/Fe] is more 
tightly bound to [Fe/H] rather than age; this might explain the reason why they 
could not detect any age-dependence in their sample solar analogs of [Fe/H]~$\simeq 0$. 
  
\section{Summary and Conclusion}

Our understanding on the galactic chemical evolution of potassium is 
still insufficient, because the observed [K/Fe] relation can not be 
well reproduced by theoretical predictions. On the observational side,
an embarrassing problem regarding galactic disk stars 
($-1 \ltsim$~[Fe/H]~$\ltsim +0.5$) is that the slopes of [K/Fe] vs. [Fe/H] 
relations derived by several previous studies are diversified and  
not consistent with each other, though the qualitative tendency (increasing 
[K/Fe] with a decrease in [Fe/H]) is the same. Besides, recent precision
abundance analysis on solar twins of widely different ages reported that
[K/Fe] is nearly constant around $\sim 0$ without showing any systematic
trend with stellar age, which is hard to understand in context of 
expected evolution of [K/Fe] with time.

With an aim to shed light to clarifying these issues, 
an extensive determination of potassium abundances was carried out 
based on the non-LTE profile-fitting analysis of the K~{\sc i} 7699 line 
applied to a large sample of 488 stars (160 late F--early K dwarfs, 
239 late G--early K nearby giants, and 89 late G--early K giants in 
the {\it Kepler} field), all belonging to the disk population 
($-1 \ltsim$~[Fe/H]~$\ltsim 0.5$), for which the spectra are available 
and the stellar parameters are already determined (stellar ages are 
known for FGK dwarfs and {\it Kepler} giants).

The main point of interest was to elucidate the observed behaviors 
of [K/Fe] in the galactic disk, where special attention was paid to 
(i) checking whether these different star groups (especially dwarfs 
vs. giants) yield consistent results with each other, (ii) clarifying 
the d[K/Fe]/d[Fe/H] gradient to compare with previous studies. and (iii) 
checking whether [K/Fe] shows any systematic trend with stellar age.

The qualitative trend turned out to be almost the same irrespective of 
the star group that [K/Fe] tends to increases with a decrease in [Fe/H]; 
especially, thick-disk stars of comparatively lower metallicity tend to 
have supersolar [K/Fe] ($>0$). However, quantitative features are 
more or less different from group to group.

The [K/Fe] vs. [Fe/H] relation of dwarf stars shows a tight tendency 
with small dispersion, and nearby giants also exhibit an almost similar 
trend to the case of dwarfs, though the dispersion is somewhat larger 
with a slightly steeper slope and a marginal offset by a few hundredths dex. 
The corresponding mean d[K/Fe]/d[Fe/H] gradient for these two groups
turned out rather shallow ($\sim -0.1$ to $-0.15$ for dwarfs and 
around $\sim -0.15$ for giants). It may be concluded from these 
results that modestly consistent [K/Fe] trends with metallicity were 
obtained for both dwarfs and giants as long as those solar neighborhood 
stars are concerned. 

As to the comparison with diversified [K/Fe] vs. [Fe/H] trends
reported so far, the d[K/Fe]/d[Fe/H] gradient ($\sim -0.1$ to $-0.15$)
resulting from this investigation turned out to be shallower 
than any of the previous studies.

The [K/Fe] values of {\it Kepler} giants are generally enhanced 
by $\ltsim 0.1$~dex (tending to be supersolar irrespective of the 
metallicity) with a considerable scatter (making the gradient of [K/Fe] 
with respect to [Fe/H] rather ambiguous). This implies that the chemical 
evolution history of these {\it Kepler} giants would have been different 
from that of solar neighborhood stars. The reason why they show so 
diversified [K/Fe] may be due to their widely scattered locations. 

Regarding the age-dependence of potassium abundance, a marginal trend 
of increasing [K/Fe] with stellar age is recognized for dwarfs, while 
any systematic tendency is not observed for {\it Kepler} giants.
The reason why Spina et al. (2016) and Reggiani et al. (2019) could not
detect any dependence upon age in the [K/Fe] values of nearby solar  
twins may be due to their specific samples (same metallicity stars of
widely different ages). It may be possible that the galactic chemical 
evolution of K is not necessarily a monotonic function of time but 
more significantly depends upon the local condition, by which [K/Fe] 
would be more tightly bound to [Fe/H] rather than age.

\bigskip

This investigation is based in part on the data collected at Subaru Telescope, 
which is operated by the National Astronomical Observatory of Japan.
 
This research has made use of the data from the European Space 
Agency (ESA) mission {\it Gaia} (https://www.cosmos.esa.int/gaia), processed 
by the {\it Gaia} Data Processing and Analysis Consortium (DPAC,
https://www.cosmos.esa.int/web/gaia/dpac/consortium). Funding for the DPAC
has been provided by national institutions, in particular the institutions
participating in the {\it Gaia} Multilateral Agreement.

\newpage

\appendix
\section*{Comparison with Reggiani et et.'s (2019) Non-LTE Calculations}

Takeda et al. (2002; cf. Appendix therein) published the non-LTE equivalent widths 
and non-LTE corrections for both the K~{\sc i} 7699 and 7665 lines calculated 
on an extensive grid of 300 models resulting from combinations of five $T_{\rm eff}$
(4500, 5000, 5500, 6000, 6500), five $\log g$ (1.0, 2.0, 3.0, 4.0, 5.0),
four [Fe/H] ($0$, $-1$, $-2$, $-3$), and three $v_{\rm t}$ (1, 2, 3) values.
The adopted atomic data for the calculations were the same as used in 
Takeda et al. (1996). For example, regarding collisions with neutral hydrogen atoms 
(generally important in non-LTE calculations for late-type stars), the rates 
computed by the conventionally used classical formula were drastically 
reduced by multiplying by an empirically determined factor of $k (= 10^{h}) = 10^{-3}$; 
and as to van der Waals damping width (which plays a significant role in calculating 
profiles or equivalent widths of strong lines), the classically evaluated damping
constant $C_{6}^{\rm classic}$ (so-called Uns\"{o}ld approximation)      
was increased by an empirically established correction of $\Delta\log C_{6} = +1.0$ 
(corresponding to multiplying the classical van der waals damping width 
$\Gamma_{6}^{\rm classic}$ by a factor of 2.5); see Takeda et al. (1996) 
for more details.

Very recently, Reggiani et al. (2019) conducted new non-LTE calculations for
neutral potassium by using the realistic atomic data achieved by modern 
physics without invoking any classical formula or empirical corrections. 
They also published non-LTE equivalent widths and non-LTE corrections
of K~{\sc i} 7699 and 7665 lines computed for a large number of models in the 
parameter range of 4000~K~$\le T_{\rm eff} \le$~8000~K, 0.5~$\le \log g \le$~5.0, 
$-5.0 \le$~[Fe/H]~$\le +0.5$, and three $v_{\rm t}$ of 1, 2, and 5~km~s$^{-1}$. 

It is interesting to check how Takeda et al.'s (2002) grids of non-LTE line strengths 
and abundance corrections are compared with those Reggiani et al.'s (2019) new calculations. 
Regarding the input K abundance, Takeda et al. (2002) adopted [K/Fe] = 
$-0.3$, 0.0, +0.3 for [Fe/H] = 0 models, while +0.2, +0.5,+0.8 
for metal-poor [Fe/H]~($<0$) models. In contrast, Reggiani et al. (2019) varied the 
K abundance 11 times as [K/Fe] = $-1.25$, $-1.00$, $-0.75$, $\cdots$, 
+1.00, and +1.25) irrespective of [Fe/H].
Therefore, in order to enable a direct comparison at the same set of input parameters, 
comparisons are made only on the cases of K~{\sc i} 7699 line with 
[Fe/H] = [K/Fe] = 0, $v_{t} = 2$~km~s$^{-1}$, and combinations of ($T_{\rm eff}$ = 
4500, 5000, 5500, 6000, 6500) and ($\log g$ = 2, 3, 4, 5).

The mutual comparisons of the non-LTE equivalent widths and non-LTE corrections
are depicted in Figures~9a and 9b, respectively.
We can see from these figures that both are in reasonable consistency, though
$W^{\rm NLTE}$ (Reggiani) tends to be somewhat lower than $W^{\rm NLTE}$ (Takeda)
when the line is very strong ($W \gtsim 500$~m\AA) and
$\Delta^{\rm NLTE}$ (Reggiani) begins to deviate from $\Delta^{\rm NLTE}$ (Takeda)
near to $\Delta^{\rm NLTE} \sim -1$ (around the largest $|\Delta^{\rm NLTE}|$).
Accordingly, it can be stated that Takeda et al.'s (2002) results based on their 
somewhat outdated calculations (upon which this study is based) are not significantly 
different from those of Reggiani et al.'s (2019) new non-LTE calculations 
using up-to-date atomic data. 


\newpage
\onecolumn

\setcounter{figure}{0}
\begin{figure*}[p]
  \begin{center}
    \FigureFile(120mm,160mm){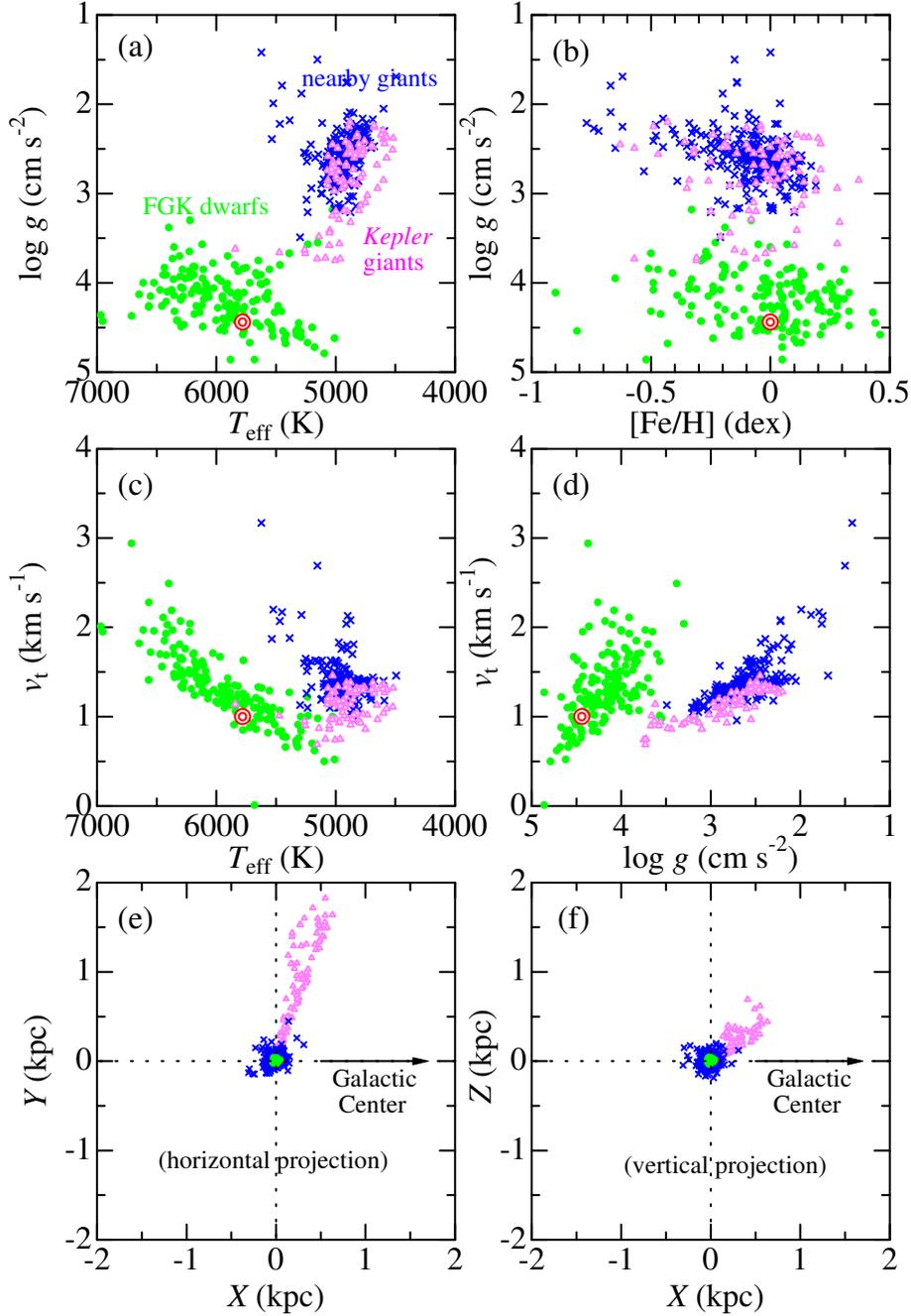}
  \end{center}
\caption{
Panels (a)--(d) show the mutual correlations of adopted atmospheric
parameters: (a) $\log g$ vs. $T_{\rm eff}$, (b) $\log g$ vs. [Fe/H],
(c) $v_{\rm t}$ vs. $T_{\rm eff}$, and (d) $v_{\rm t}$ vs. $\log g$.
The lower two panels (e) and (d) are the cross sectional diagrams
projected onto the $Z = 0$ plane (galactic plane passing the Sun) and 
the $Y = 0$ plane (plane perpendicular to the galactic plane, 
passing the Sun and the galactic center), respectively,
which illustrate the locations of the program stars in the Galaxy 
(Sun at the origin of $X = Y = Z = 0$).
Three groups of program stars are distinguished by different symbols:
green circles --- 160 FGK dwarfs, blue crosses --- 239 nearby giants,
and pink triangles --- 89 {\it Kepler} giants. The Sun is indicated by 
the red double circle.    
}
\end{figure*}

\setcounter{figure}{1}
\begin{figure*}[p]
  \begin{center}
    \FigureFile(140mm,170mm){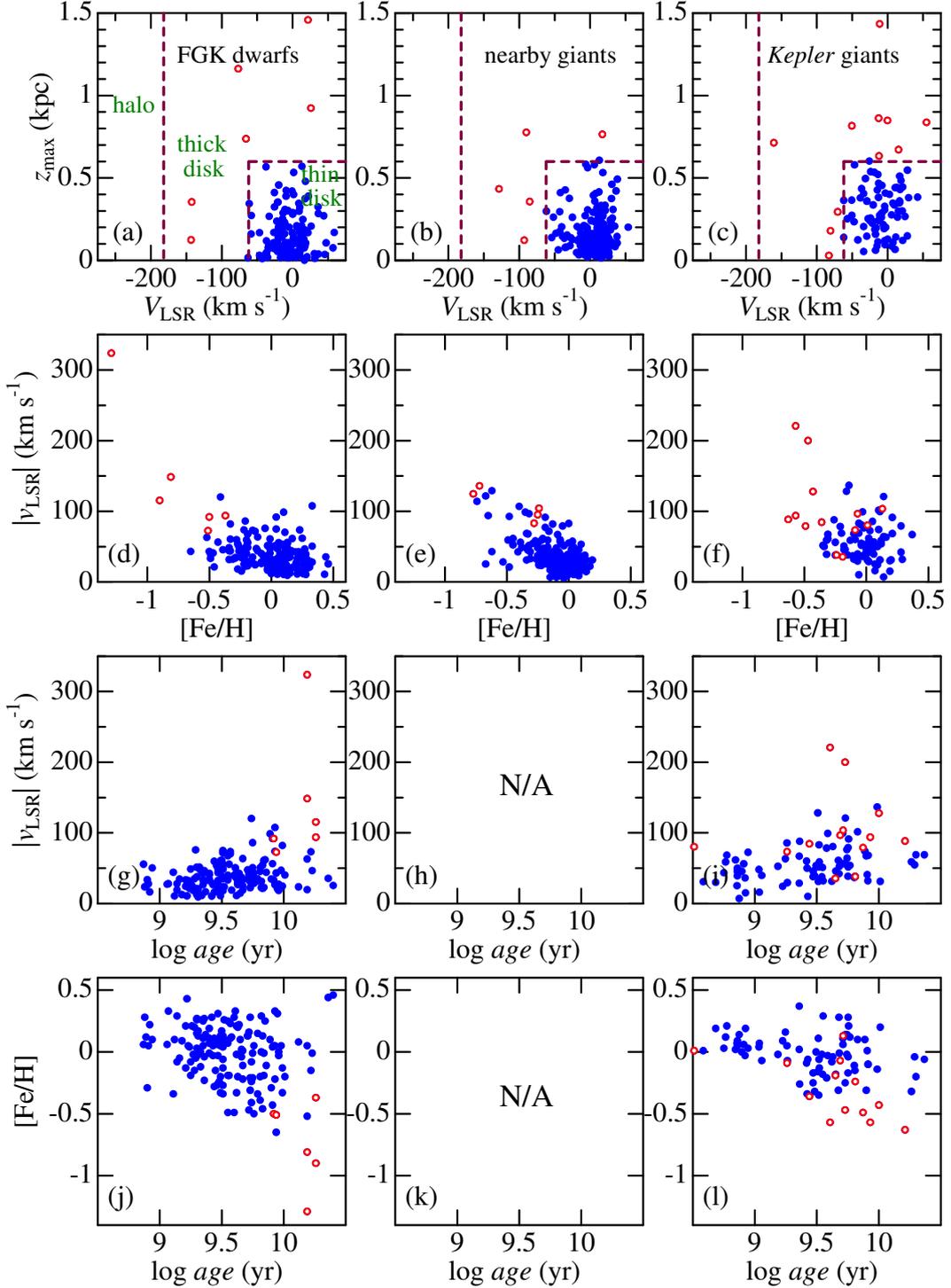}
  \end{center}
\caption{
In the top (1st) row are shown the correlation diagrams 
between the maximum separation from the galactic plane ($z_{\rm max}$)
and the rotation velocity component relative to LSR ($V_{\rm LSR}$), 
which may be used for classifying the stellar population (the boundaries 
are indicated by the dashed lines; cf. Ibukiyama \& Arimoto 2002).
The extents of space velocities relative to LSR
[$|v_{\rm LSR}| \equiv (U_{\rm LSR}^{2} + V_{\rm LSR}^{2} + 
W_{\rm LSR}^{2})^{1/2}$] are plotted against [Fe/H] and $age$ 
in the 2nd and 3rd rows, respectively.
The $age$ vs. metallicity ([Fe/H]) relations are depicted in the 
bottom (4th) panels. 
The left, middle, and right panels correspond to FGK dwarfs, 
nearby giants, and {\it Kepler} giants, respectively.
Stars of thin-disk and thick-disk populations are indicated by (blue) 
filled and (red) open symbols, respectively. Note that, in panel (c), 
one star (KIC~2714397) of thick-disk population ($V_{\rm LSR} = -181$~km~s$^{-1}$, 
$z_{\rm max} = 3.17$~kpc) is outside of the plot range. 
}
\end{figure*}

\setcounter{figure}{2}
\begin{figure*}[p]
  \begin{center}
    \FigureFile(160mm,200mm){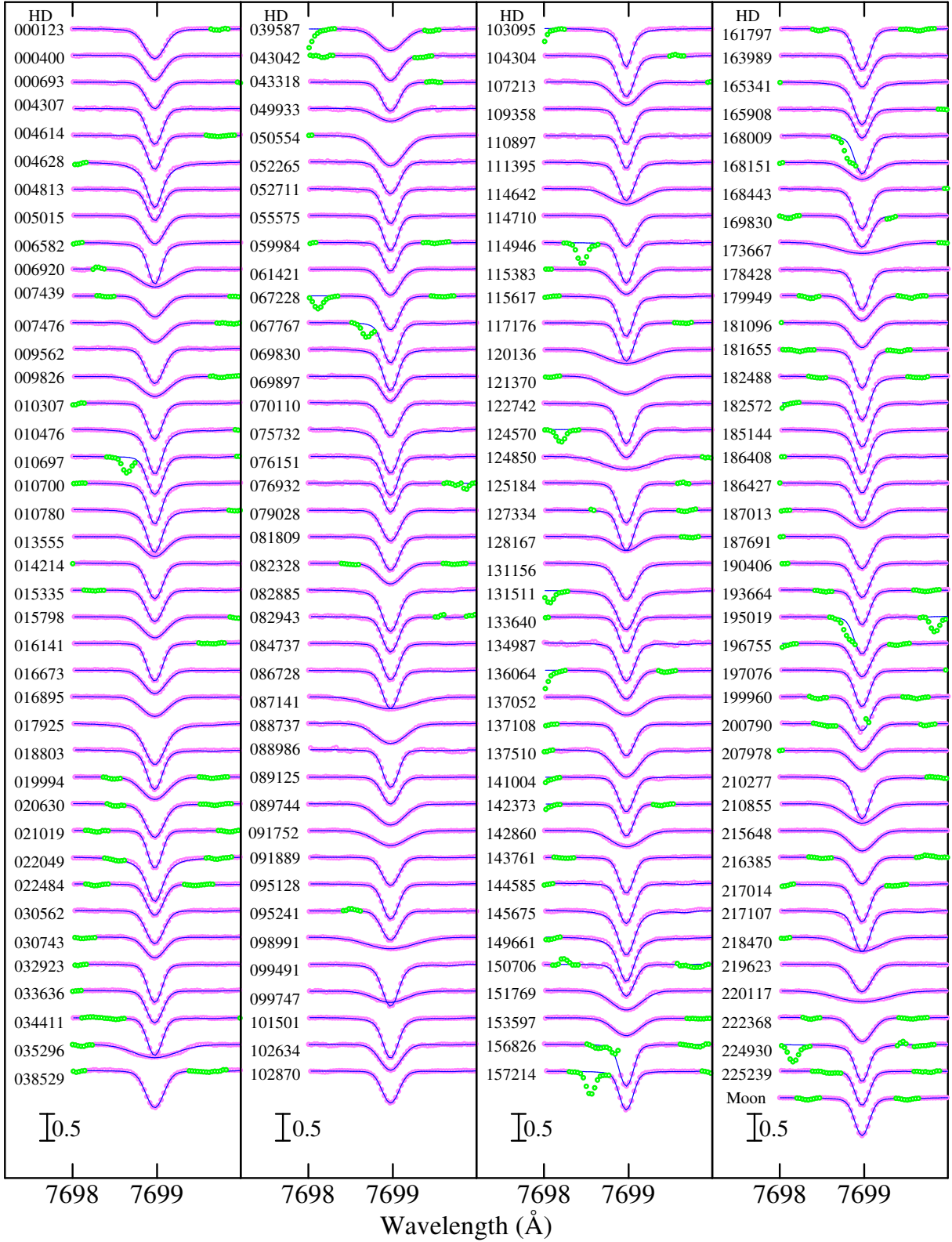}
  \end{center}
\caption{
Synthetic spectrum fitting for 160 FGK dwarfs/subgiants (and the Sun/Moon) 
in the region comprising the K~{\sc i} 7699 line. The best-fit theoretical 
spectra are shown by blue solid lines, and the observed data are plotted
by pink open circles (while those masked/disregarded in the fitting 
are highlighted in green). A vertical offset of 0.5 (in terms of the 
normalized flux with respect to the continuum) is applied to each spectrum
relative to the adjacent ones. Each of the spectra are arranged 
in the increasing order of star number (indicated on the left to each 
spectrum). The wavelength scale of each spectrum is adjusted to 
the laboratory frame.
}
\end{figure*}

\setcounter{figure}{3}
\begin{figure*}[p]
  \begin{center}
    \FigureFile(160mm,200mm){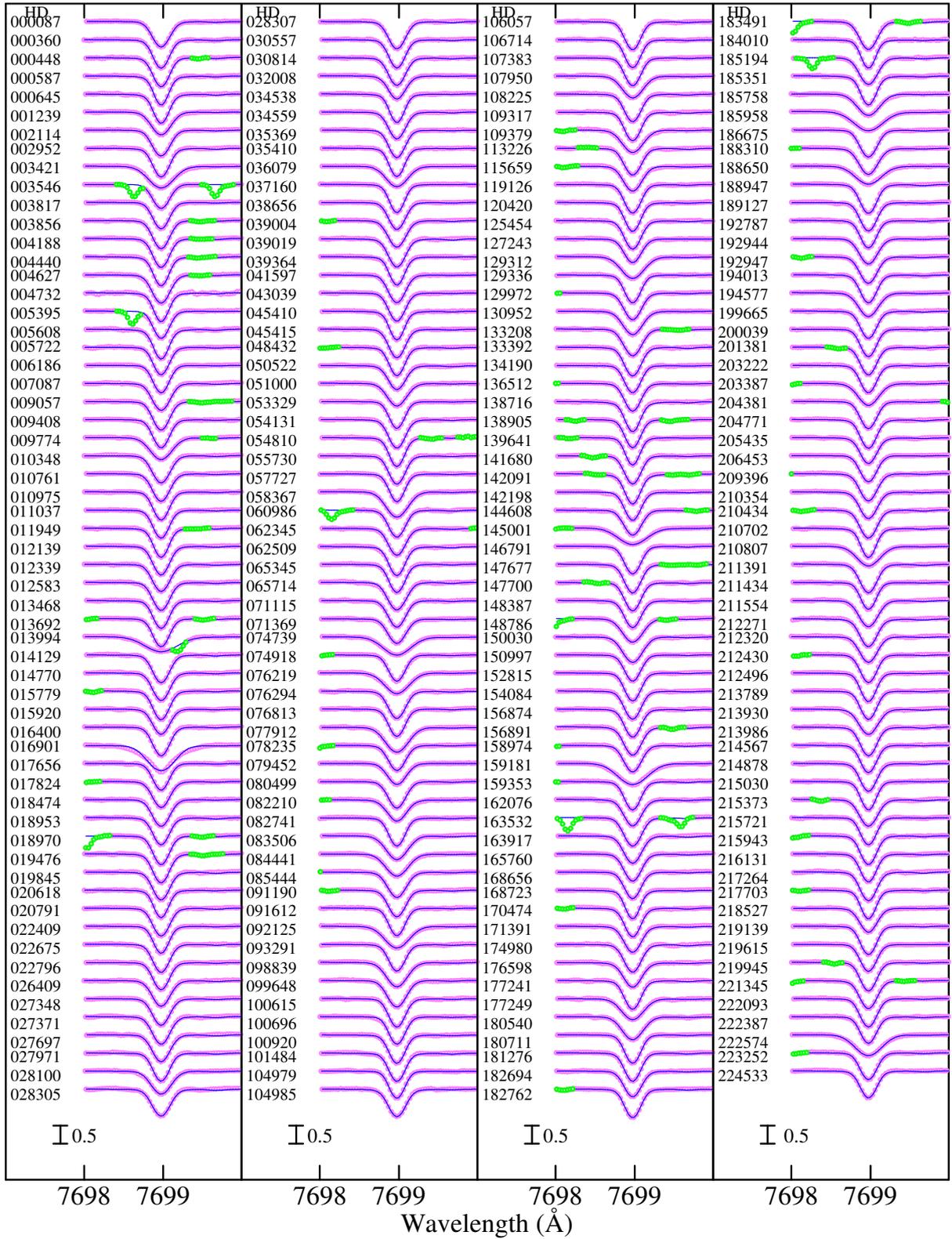}
  \end{center}
\caption{
Synthetic spectrum fitting for 239 nearby red giants.
Otherwise, the same as in Figure~3.
}
\end{figure*}

\setcounter{figure}{4}
\begin{figure*}[p]
  \begin{center}
    \FigureFile(120mm,160mm){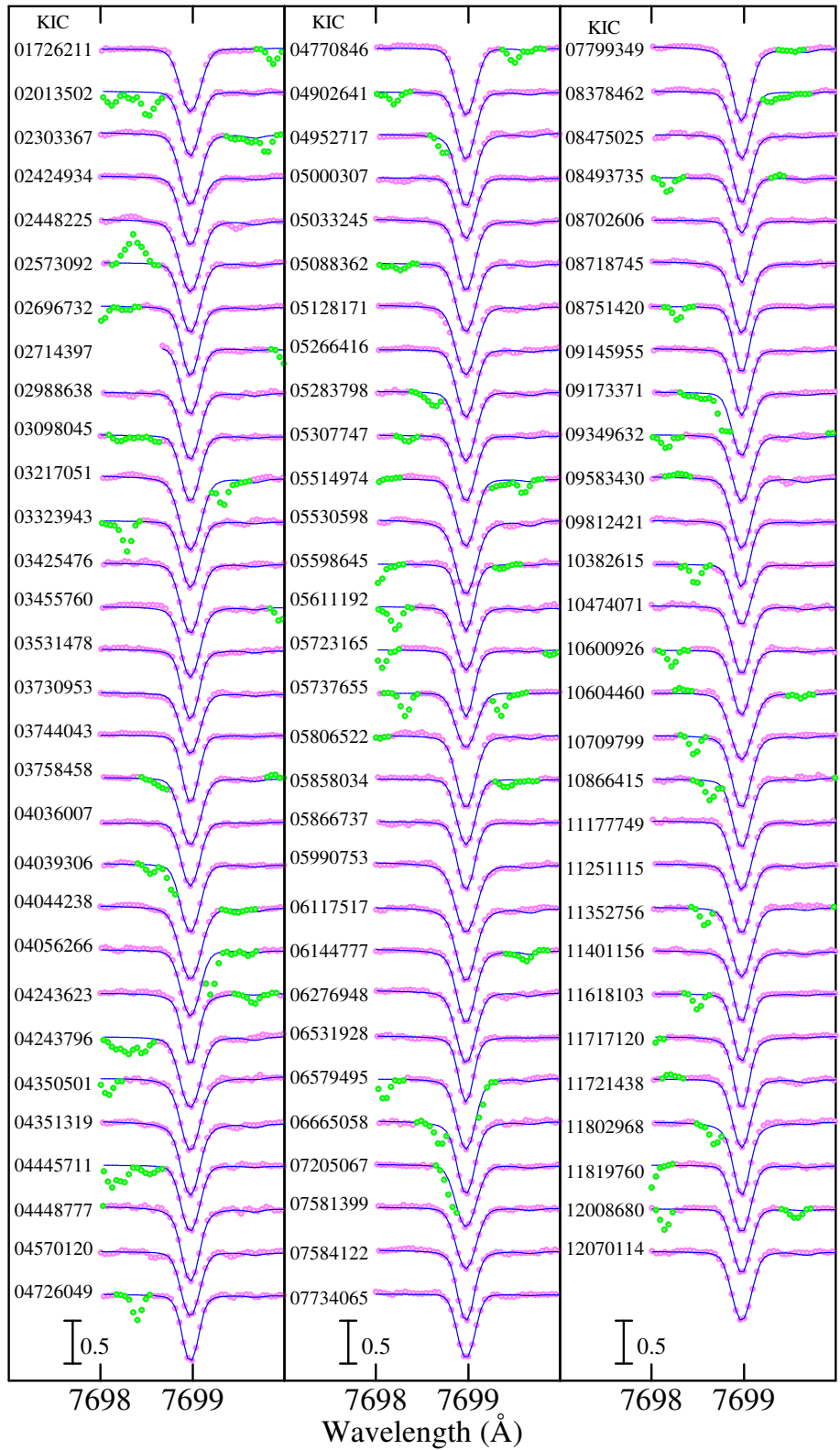}
  \end{center}
\caption{
Synthetic spectrum fitting for 89 giants in the {\it Kepler} field.
Otherwise, the same as in Figure~3.
}
\end{figure*}

\setcounter{figure}{5}
\begin{figure*}[p]
  \begin{center}
    \FigureFile(120mm,160mm){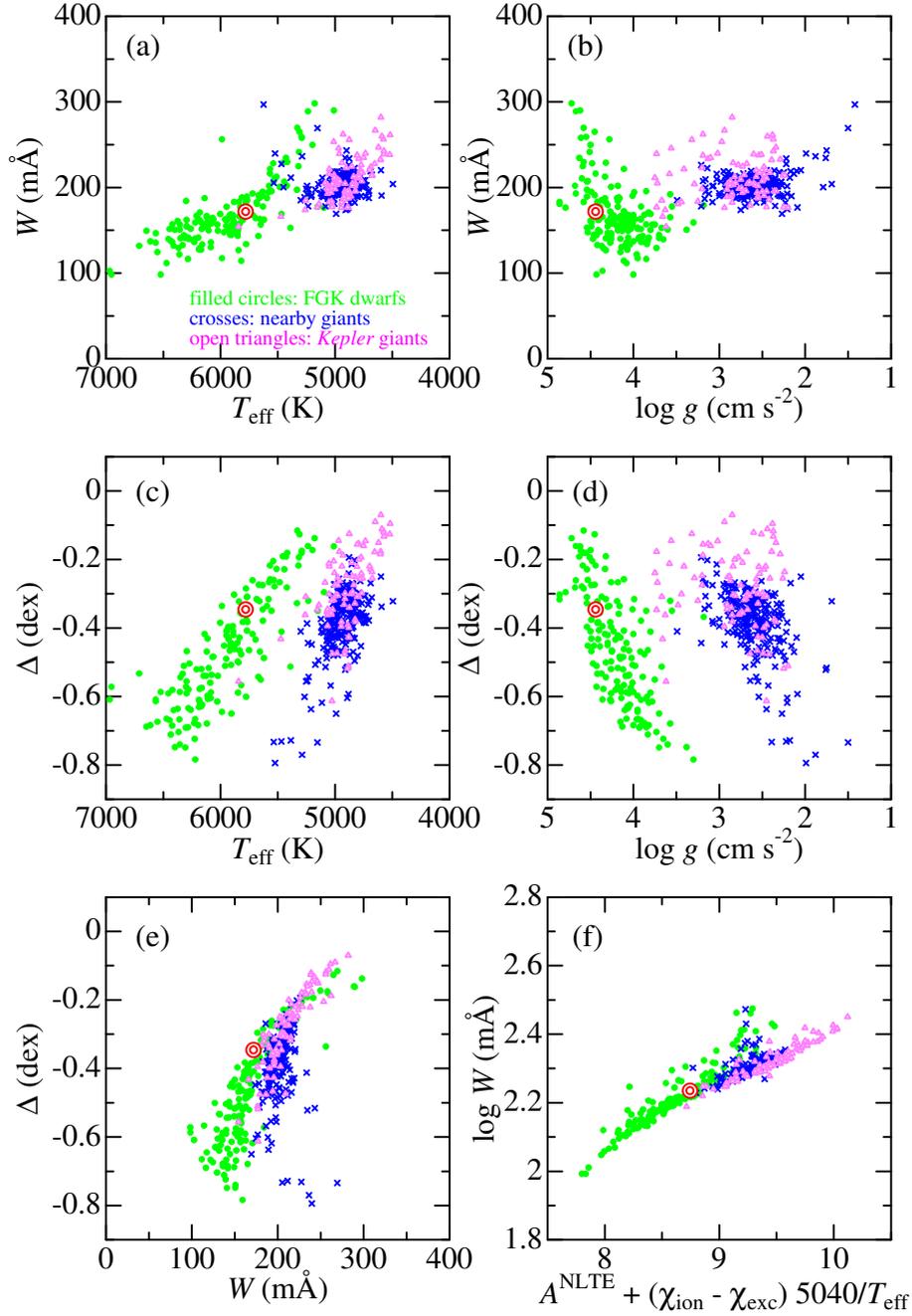}
  \end{center}
\caption{
Equivalent widths ($W$) of the K~{\sc i} 7699 line are plotted against 
$T_{\rm eff}$ and $\log g$ in panels (a) and (b), while the dependences
of non-LTE corrections ($\Delta$) upon $T_{\rm eff}$, $\log g$, and
$W$ are shown in panels (c), (d), and (e), respectively.
In panel (f) are plotted the $\log W$ values against  
$A^{\rm NLTE}$ + $(\chi_{\rm ion} - \chi_{\rm exc})5040/T_{\rm eff}$
(abscissa of curve of growth), where $A^{\rm NLTE}$ is the non-LTE abundance 
derived for each star and $\chi_{\rm ion}$/$\chi_{\rm exc}$ is the 
ionization/excitation potential (in eV). The same meanings of the 
symbols as in Figure~1. 
}
\end{figure*}

\setcounter{figure}{6}
\begin{figure*}[p]
  \begin{center}
    \FigureFile(150mm,180mm){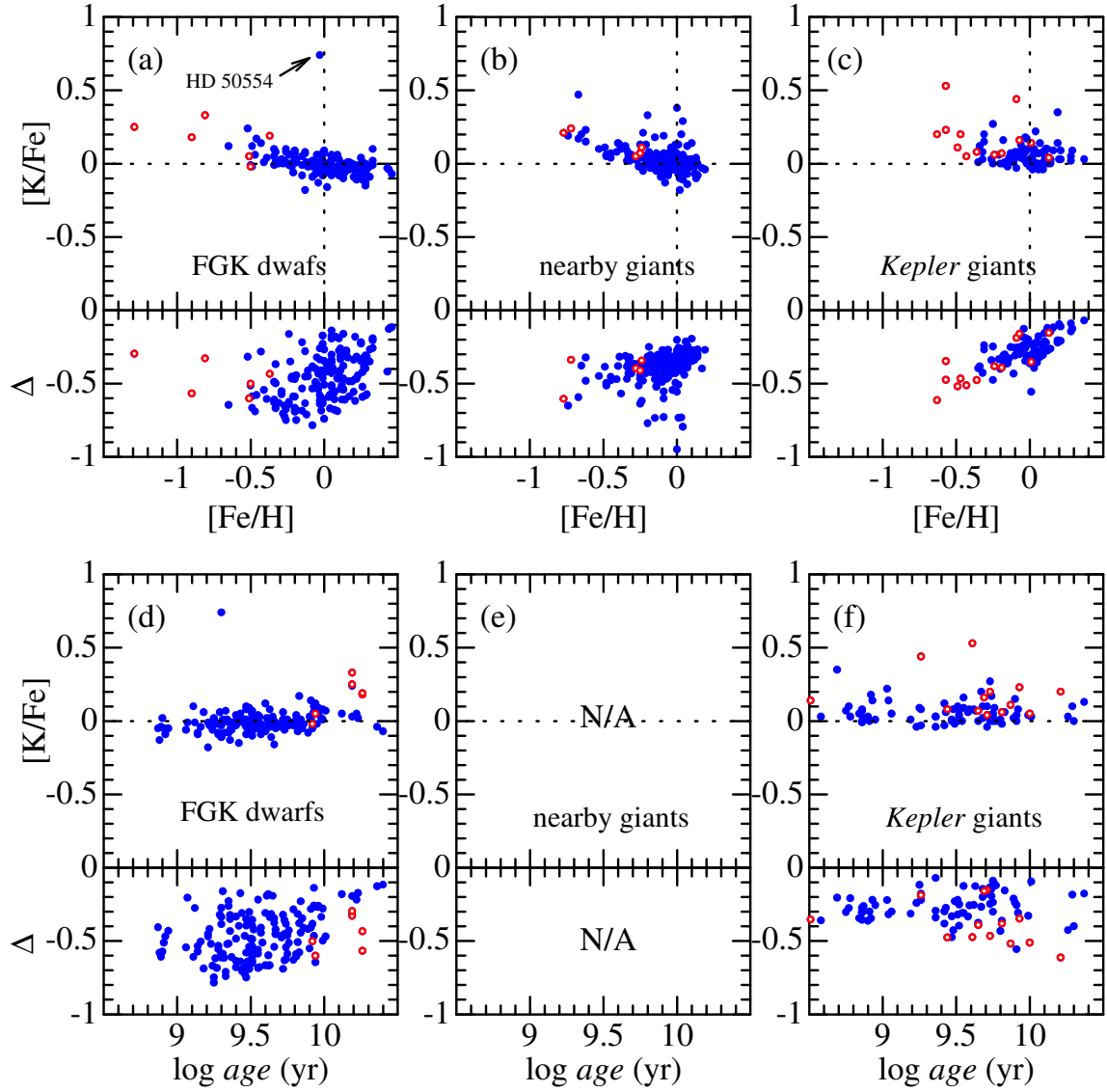}
  \end{center}
\caption{
The values of [K/Fe] (non-LTE K-to-Fe abundance ratio) and $\Delta$ 
(non-LTE correction) derived for each star are plotted against [Fe/H] 
(upper panels) and $\log age$ (lower panels), respectively.  
As in Figure~2, the left, middle, and right panels correspond 
to FGK dwarfs, nearby giants, and {\it Kepler} giants, respectively.
Likewise, the same meanings of the symbols hold as in Figure~2.
}
\end{figure*}

\setcounter{figure}{7}
\begin{figure*}[p]
  \begin{center}
    \FigureFile(80mm,120mm){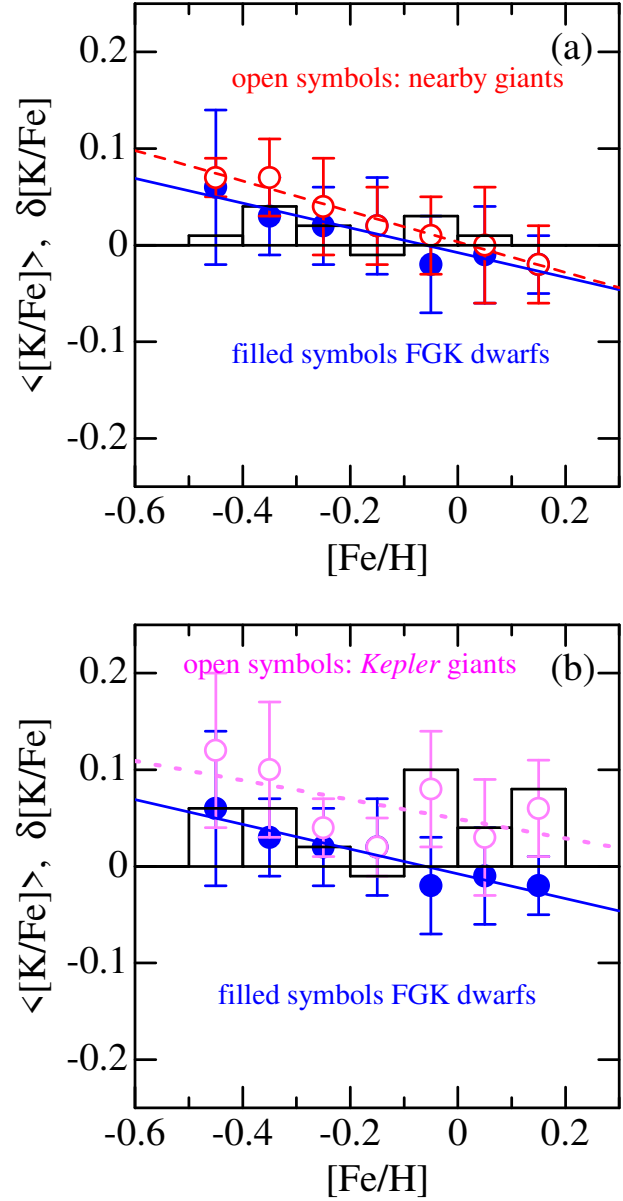}
  \end{center}
\caption{
The mean $\langle$[K/Fe]$\rangle$  values calculated at each metallicity 
group (0.1~dex bin within $-0.5 \le$~[Fe/H]~$\le +0.2$) are illustrated by 
symbols (attached error bars denote standard deviations), where filled 
and open circles correspond to dwarfs and giants, respectively. 
The bar graphs represent the giants$-$dwarfs differences 
($\langle$[K/Fe]$\rangle_{\rm giants} - \langle$[K/Fe]$\rangle_{\rm dwarfs}$). 
The upper panel (a) shows the comparison of nearby giants vs. FGK dwarfs
while the lower panel (b) is for {\it Kepler} giants vs. FGK dwarfs.
The linear regression lines determined from these $\langle$[K/Fe]$\rangle$
values (cf. Section~5.2) are also drawn by solid, dashed, and dotted lines
for FGK dwarfs, nearby giants, and {\it Kepler} giants, respectively. 
}
\end{figure*}

\setcounter{figure}{8}
\begin{figure*}[p]
  \begin{center}
    \FigureFile(80mm,120mm){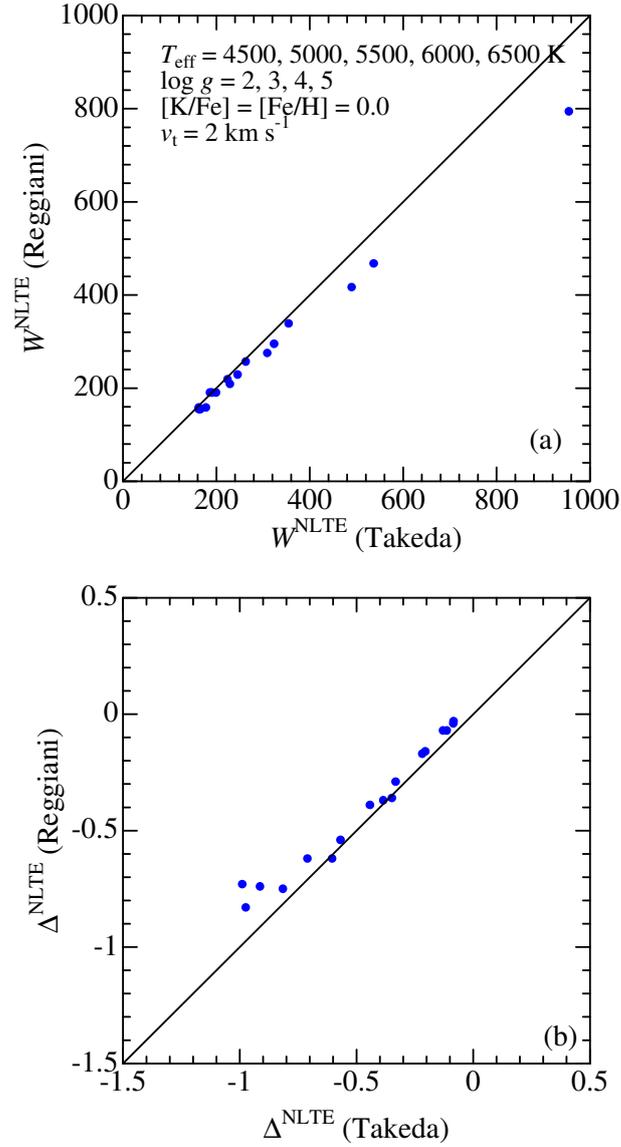}
  \end{center}
\caption{
Comparison of the non-LTE equivalent widths ($W^{\rm NLTE}$: upper panel) 
and non-LTE corrections ($\Delta^{\rm NLTE}$: lower panel).
In the ordinate are the values taken from Reggiani et al.'s (2019) non-LTE grid 
computed for the K~{\sc i}~7698.974 line by using the recent up-to-date atomic data, 
while those published by Takeda et al. (2002) are in the abscissa. 
Shown here are the cases of [K/Fe] = [Fe/H] = 0 and $v_{\rm t}$ = 2~km~s$^{-1}$  
with the combinations of ($T_{\rm eff}$ = 4500, 5000, 5500, 6000, 6500~K)
and ($\log g$ = 2, 3, 4, and 5), for which direct comparisons are possible.
}
\end{figure*}

\end{document}